\documentclass[11pt]{article}
\usepackage{epsfig}
\usepackage{oldgerm}
\input{amssym}
\def\be{\begin{equation}}
\def\ee{\end{equation}}

\newsavebox{\PSLASH}
\sbox{\PSLASH}{$p$\hspace{-1.8mm}/}

\begin{document}
\title{Continuous Abelian Sandpile Model in Two Dimensional Lattice}
\author{N. Azimi-Tafreshi, E. Lotfi, S. Moghimi-Araghi\footnote{e-mail: samanimi@sharif.edu}\\
Department of Physics, Sharif University of Technology,\\ Tehran,
P.O.Box: 11365-9161, Iran} \date{} \maketitle

\begin{abstract}
We define a new version of sandpile model which is very similar to
Abelian Sandpile Model (ASM), but the height variables are
continuous ones. With the toppling rule we define in our model, we
show that the model can be mapped to ASM, so the general properties
of the two models are identical. Yet the new model allows us to
investigate some problems such as effect of very small mass on the
height probabilities, different boundary conditions, etc.

  \vspace{5mm}%
\textit{PACS}: 64.60.av, 45.70.Cc, 05.65 +b, 
\newline \textit{Keywords}:
Self-Organized Criticality, Sandpile models.
\end{abstract}

\section{Introduction}

The concept of self-organized criticality (SOC) was first introduced
by Bak, Tang and Wiesenfeld \cite{BTW} through a simple model. This
concept is believed to be the underlying reason for the scaling laws
seen in a number of natural phenomena \cite{Jensen}.  Their model,
which was called Abelian Sandpil Model (ASM) is still the simplest,
most studied model of SOC, in which many analytical results has been
derived. For a good review see ref. \cite{DharRev}.

Many exact results are derived in this theory. The first analytical
calculation, which paved the road for other analytical results, was
done by Dhar \cite{Dhar}. In his paper, Dhar computed the number of
recurrent configuration and showed that all occur with equal
probability. Later, in \cite{MajDharJPhysA}, Majumdar and Dhar
calculated the probabilities of occurrence of some specific
clusters, known as Weakly Allowed Clusters (WAC's). The simplest of
these clusters is one-site height one cluster. The probabilities of
other one-site clusters with height above 1 was computed in ref.
\cite{Prizh}. There are many other analytical results, among them
one can mention the results on boundary correlations of height
variables and effect of boundary conditions
\cite{Ivashkevich94,Jeng05-1,Jeng05-2,Jeng04,Ruelle02,Ruelle07}, on
presence of dissipation in the model \cite{Jeng05-2,Jeng04,PirRu04},
on field theoretical approaches \cite{MaRu,Jeng05-2,MRR,JPR,MN}, on
finite size corrections \cite{Ruelle02,MajDharJPhysA} and many other
results\cite{DharRev}.

In ASM, the height variables take only integer values. There are
some other models that in which the height variables can be real
\cite{Zhang,Gabrielov}. In Zhang model, when a site topples, all of
its energy (height) is distributed among it neighbors and nothing is
left for the original site. The dynamics does not have the Abelian
property that the ordinary sandpile model has. Therefore, few
analytical results are derived in this model \cite{Redig}. On the
other hand the model introduced by Gabrielov, called Abelian
Avalanche Model (AA Model), is Abelian and more or less resembles
the original ASM. The model is defined on a general graph and many
general properties are investigated. In this paper we introduce a
model similar AA model, defined on a square lattice and investigate
some more specific properties. This model, in which the height
variables can adopt any (positive) real numbers, allows us to
investigate some yet unanswered question in the original ASM, such
as the effect of dissipation on all height probabilities.

In this paper we first introduce our model, and show that it can be
mapped to the usual ASM. Using analytical results and simulations,
we derive some properties of the model. Next we introduce
dissipation to the model and see what will be its effect
probabilities of height variables.  Then we investigate the model in
presence of some boundary conditions.

\section{Continuous Abelian Sandpile Model }

 Consider a $L\times L$
square lattice. To each site $u=(i,j)$ a {\it continuous} height
variable, $h(u)$ is assigned, where without loss of generality we
assume these variables are in [0,1). The evolution rules consist of
the following rules: 1) At each time step, a site is selected
randomly and an amount of sand is added to it and other height
variables are unchanged. The amount of sand added to the site is a
random real number in the interval $[p,q]\in[0,1)$. The distribution
function could be any function, but for simplicity we take it to be
uniform on the interval $[p,q]$. If height remains below one, the
new configuration is stable and we go to the next time step. 2) If
height of the sand at that site becomes equal to or greater than
one, the site becomes unstable and topples in the following way: it
gives an amount of sand with height 1/4 to either of its neighbors.
In other words $h(v)\rightarrow h(u)-\Delta^C(u,v)$ for all $v$,
where $\Delta^C(u,v)$ is the toppling matrix of the CASM and is
defined as
\begin{eqnarray}
\Delta^C (u,v) \left\{
\begin{array}{cc}
1 & u=v\\
-1/4 &  |u-v|=1\\
0 & {\rm otherwise}
\end{array}
\right.
\end{eqnarray}

So if the height of a site is $1+\epsilon$, after toppling its
height will be $\epsilon$ and the heights of neighbors will be added
by the amount 1/4. As a result of this toppling, some of the
neighbors may become unstable and an avalanche may occur. Note that
in the Zhang model, all the energy (height) of the site is
distributed to the neighbors and the site keeps no energy.

It is clear that the model, which we call it Continuous Abelian
Sandpile Model (CASM), has the same sprite as of ASM. We show in
this paper that this model can be mapped to ASM and hence has the
very same property of it; that is, it shows self-organized
criticality and some of its properties could be found analytically.
This model has some similarities to the one introduced in \cite{GLJ}
and \cite{Tsuchia}. In the paper by Ghaffare {\it et al.} , the
problem is investigated using mean field approach and the
distribution of avalanche sizes is discussed using simulation. In
the paper by Tsuchia and Katori, the height variables are still
integer numbers, but if we take a large number for the threshold
value, it will approach to our model.
 They have considered the avalanche size distribution and
correlation functions when dissipation is present in their model.
Here we investigate the model at and off criticality and consider
some boundary conditions and some other variants of the model.
\begin{figure}[t]
\begin{picture}(200,200)(0,0)
\includegraphics{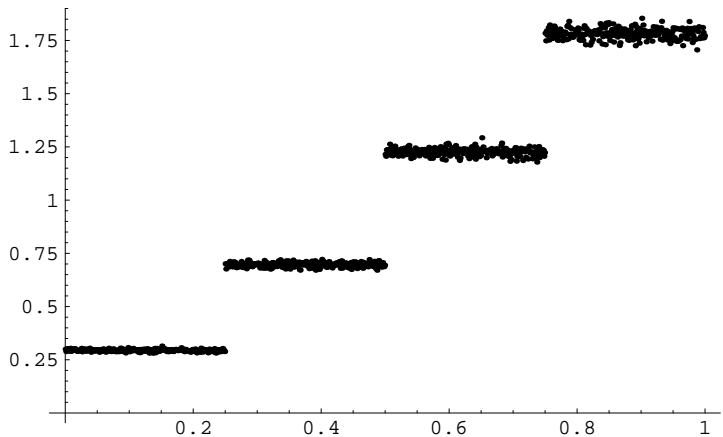}
\end{picture}
\caption{The probability density profile of height variables}
\end{figure}

A simulation is done and the probability distribution is found to be
like Fig. 1. The result is more or less the same as the one obtained
in \cite{Vazquez}. As you see, the probability distribution has four
plateaus, each have a length equal to 1/4, but their heights
increase step by step. It is easy to find the probability of finding
a site at each of these plateaus. Labeling the plateaus by numbers
1, 2, 3 and 4, it is found that $p(1)=0.0736\pm 0.00008$,
$p(2)=0.1741\pm 0.0004$, $p(3)=0.3064\pm 0.0007$, $p(4)=0.4456\pm
0.001$, where $p(k)$ is the probability of being in $k$-th plateau.
These probabilities are just the same as the ones in ASM with $p(k)$
being the probability of finding a site with height $k$. This brings
to mind that there should be a close relation between the two
models. In this section we prove two propositions:

{\bf Proposition 1:} There is a (many to one) mapping from
configurations in CASM to configurations in ASM, which preserves the
dynamics. That is, it does not matter that you first apply the
dynamics in CASM and then map it to ASM or vice versa.

{\bf Proposition 2:} The probability distribution is piecewise
constant.

To prove proposition 1, we define the map ${\mathcal M}$ from
configuration space of CASM (${\mathcal C}$) to configuration space
of ASM (${\mathcal A}$). Take $c\in{\mathcal C}$ and $a\in{\mathcal
A}$. Let's call the height variable at site $u$ in configurations
$c$ and $a$, $h^C_u$ and $h^A_u$. The action of ${\mathcal M}$ on
$c$ is $a$ if
\begin{equation}\label{mapdef}
h^A_u=f(h^C_u),
\end{equation}
where \be\label{fdef} f(x)=[4 x]+1. \ee This means that the $k$-th
1/4 goes to $k$.

Now it is easy to see that the map ${\mathcal M}$ preserves the
dynamics of the theory. The first rule of evolution is to add sand.
If the value of added sand in CASM takes the height of a site to $n$
higher plateau, it is equivalent to adding $n$ sand grains to the
site. Although it does not matter, we may assume $q \leq 1/4$ so
that at each time step, at most one sand grain is added to the site.
The other evolution rule is the toppling rule. Calling $T^C_u$ and
$T^A_u$ the toppling operators that only topple the site $u$ in CASM
and ASM respectively, we would like to show if $a={\mathcal M} c$
then $T^A_u a= {\mathcal M} T^C_u c$, that is the toppling rule and
the mapping commute, hence the map preserves the dynamic. We should
just note that the toppling matrix of ASM, $\Delta^A$, is 4 times
the toppling matrix of CASM and all of its elements are integer. So
the height variable of the configuration $ {\mathcal M} T^C_u c$ at
site $v$ will be $[4 h^C_v - \Delta^C_{u,v}]+1=[4 h^C_v]+1 -
\Delta^A_{u,v}= h^A_v- \Delta^A_{u,v}$ which is the height variable
of $T^A_u a$ at the same site. So the first proposition is proved.

The proof of second proposition is a bit lengthier. To begin, we
define the probability distribution function $P(h)dh$ which is the
probability of finding a site with height in the interval
$[h,h+dh]$. Also we define the aggregate probability distribution
function as \be P_I(\epsilon)=P\left(\epsilon\right)+P\left(\frac 1
4 +\epsilon\right)+P\left(\frac 1 2 +\epsilon\right)+P\left(\frac 3
4 +\epsilon\right)+P(1 +\epsilon), \ee where $0\leq \epsilon<1/4$.
Note that although in the stable states we have $P(1+\epsilon)$=0,
but we have added such term in this definition. It is easy to prove
that the toppling process does not change the aggregate probability
function, since in a toppling process, the height of a site either
increases by the amount 1/4 or decreases by the amount 1, that is if
it has been in the set $\{\epsilon,
1/4+\epsilon,1/2+\epsilon,3/4+\epsilon,1+\epsilon\}$ it will remain
in the same set. So the profile of the aggregate probability
function is not altered under toppling.

Now we consider the effect of the other evolution rule, the rule of
adding sand, on the aggregate probability function. As you see we
have somehow identified the heights $k/4 +\epsilon$ where $k$ is an
integer. As a result we may consider the whole interval of possible
heights as a circle with perimeter 1/4 (see Fig. 2a). Adding height
$\delta h$ to a site takes a point on this circle to a point which
is away from it with the amount $\delta h$, no matter where the
starting point is. This means that after sufficient time that the
system reaches stationary state the probability function on the
circle is uniform, that is the aggregate probability function is
uniform.

\begin{figure}
\begin{picture}(100,100)(0,0)
\includegraphics{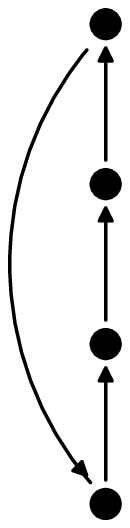} \includegraphics{levels.eps} \includegraphics{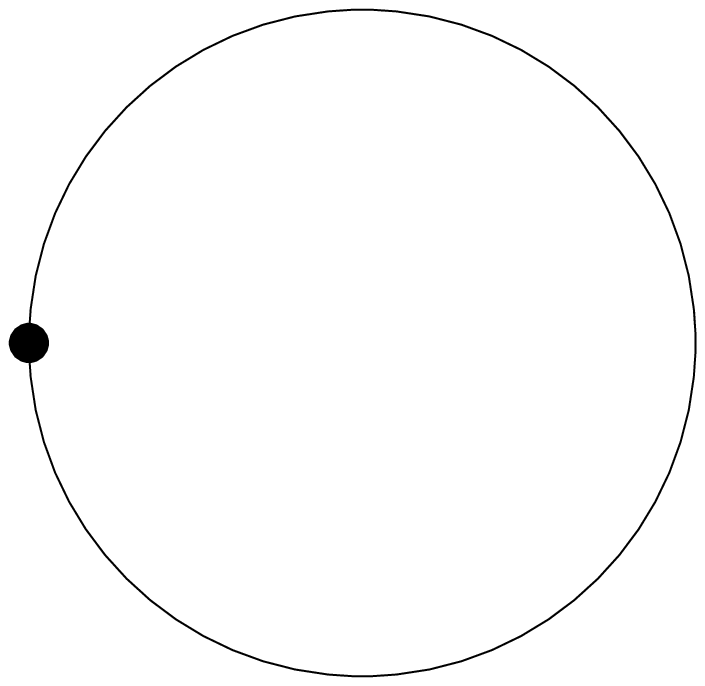}
\put(305,15){$\epsilon_1$}\put(365,15){$\epsilon_2$}\put(20,55){1/4,
1/2, 3/4, 1}\put(140,-5){(a)}\put(335,-5){(b)}
\end{picture}
\caption{(a) The aggregate probability could be defined of a circle
with perimeter 1/4. (b) The effect of toppling on relative height
probabilities for different $\epsilon$'s.}
\end{figure}

In the end we have to show that the relative probabilities of $p(k/4
\epsilon)$ are independent of $\epsilon$. The toppling process may
alter these relative probabilities. As shown in Fig. 2b, the effect
of the toppling process is the same for different $\epsilon$'s. This
is due to the fact that the transition probabilities are independent
of $\epsilon$, as all of them are mapped to the same configuration
in ASM. Therefore the relative probabilities will be the same if we
let the system evolve under many topplings and reach stationary
state. This completes the proof of the second proposition.

From these two propositions, one can determine the density in the
configuration in the stationary state. Let's call the subset of all
configurations in CASM that are mapped to the specific recurrent
configuration $a$ in $\mathcal{A}$ by $C_a$. By proposition one, it
is clear that the probability of being in $C_{a_1}$ is equal to the
probability of being in $C_{a_2}$, where $a_1$ and $a_2$ are two
recurrent configurations in ASM. By proposition two, one deduces
that within a set like $C_a$, the probability distribution is
uniform, so it is easily seen that in steady state he density in the
configuration state is uniformly distributed in the recurrent
subspace.

All of this could be seen in through a naive reasoning. Consider the
same square lattice, but assume that there are $N/4$ connections
between each two neighbors. Through any connection, only
$\delta=1/N$ grain of sand can be transferred. This means that when
a toppling occurs, the height of neighbors is increased by the
amount 1/4. As you can easily see, if we take $N=4$, apart from a
factor of 4, we arrive at the usual ASM. However, if we take the
limit of $N\rightarrow \infty$ we will have the CASM. It is easy to
see that most of the properties of this model, are the same. Also
neglecting dissipation, the model introduced in \cite{Tsuchia} is a
special version of this model. In the next section we introduce some
variants of the model that may have dissipation.

\section{Dissipative CASM}

In usual models, the dissipation parameter, which is usually called
mass, can have only integer values, though some simulations with
continuous height variables are done \cite{GLJ,Manna} and using
analytical results many aspects of massive ASM, with arbitrary
continuous mass has been derived \cite{MaRu,Jeng05-1}. To add
dissipation to the model introduced in the previous section, one can
increase the threshold height to $1+t$ instead of one, where $t$ is
a positive real number. This means that if the height of the site
$u$ becomes equal to or greater than $1+t$ then
\begin{figure}[t]
\begin{picture}(200,240)(0,-20)
\includegraphics{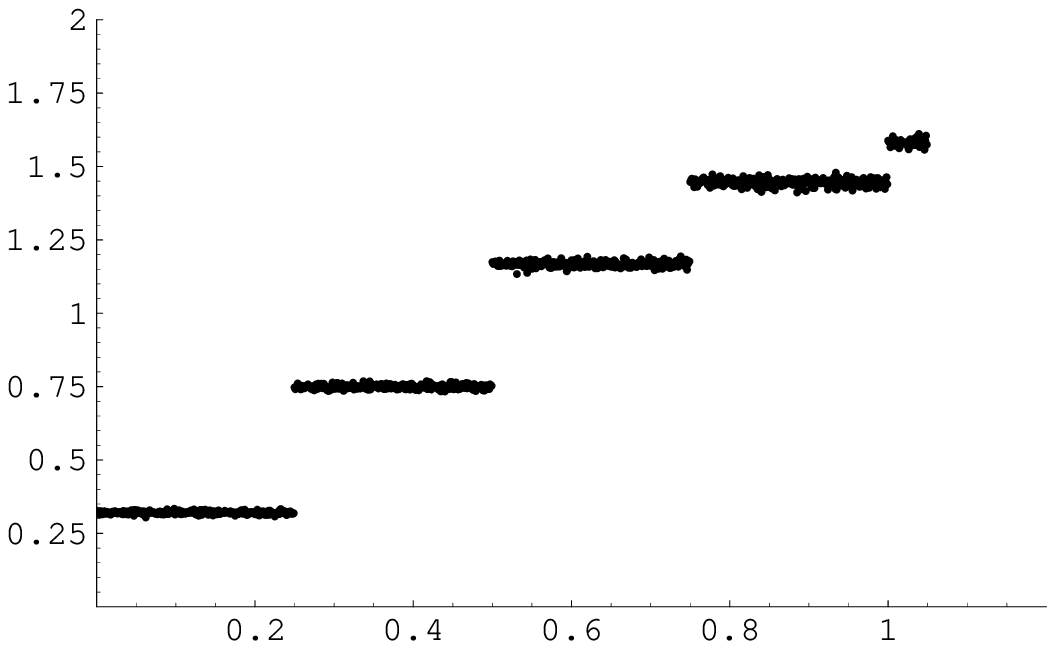}\put(130,106){(a)} \includegraphics{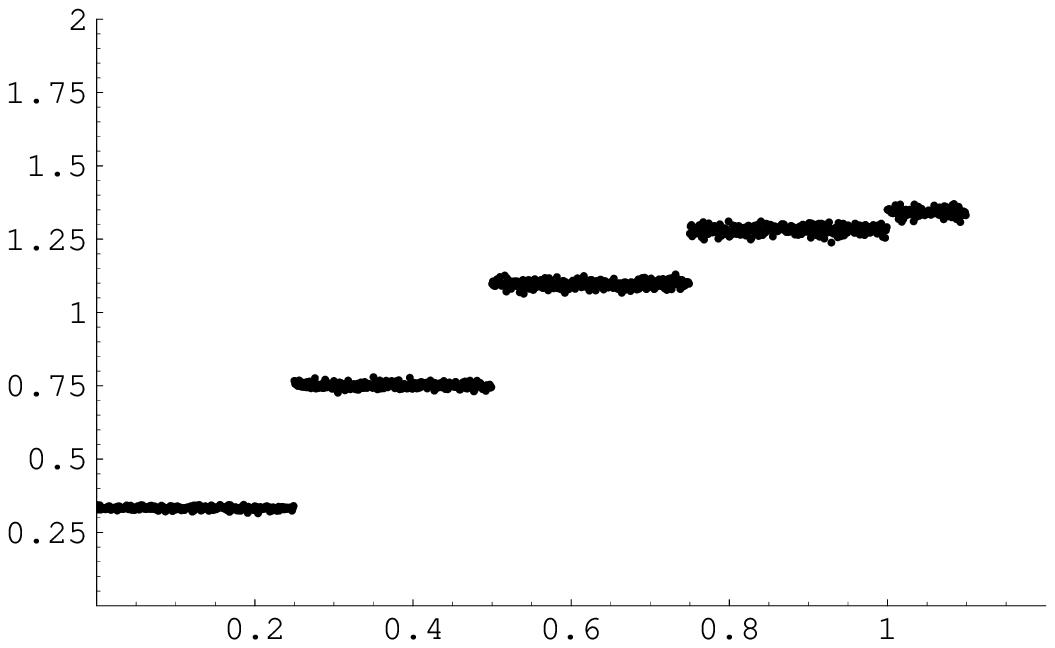}\put(310,106){(b)}
\includegraphics{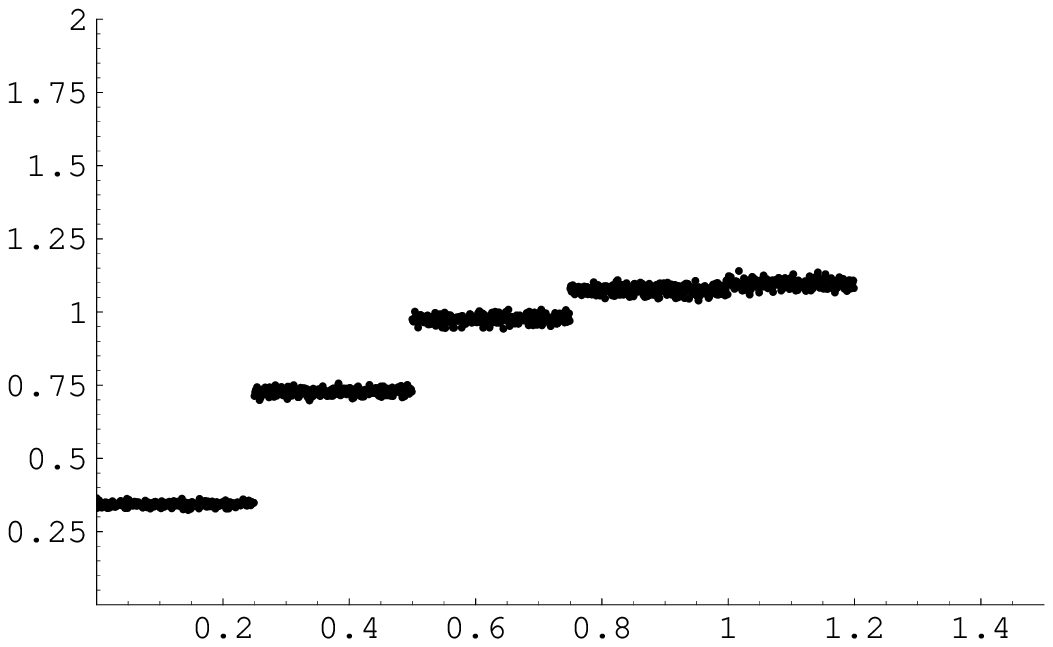}\put(130,-15){(c)} \includegraphics{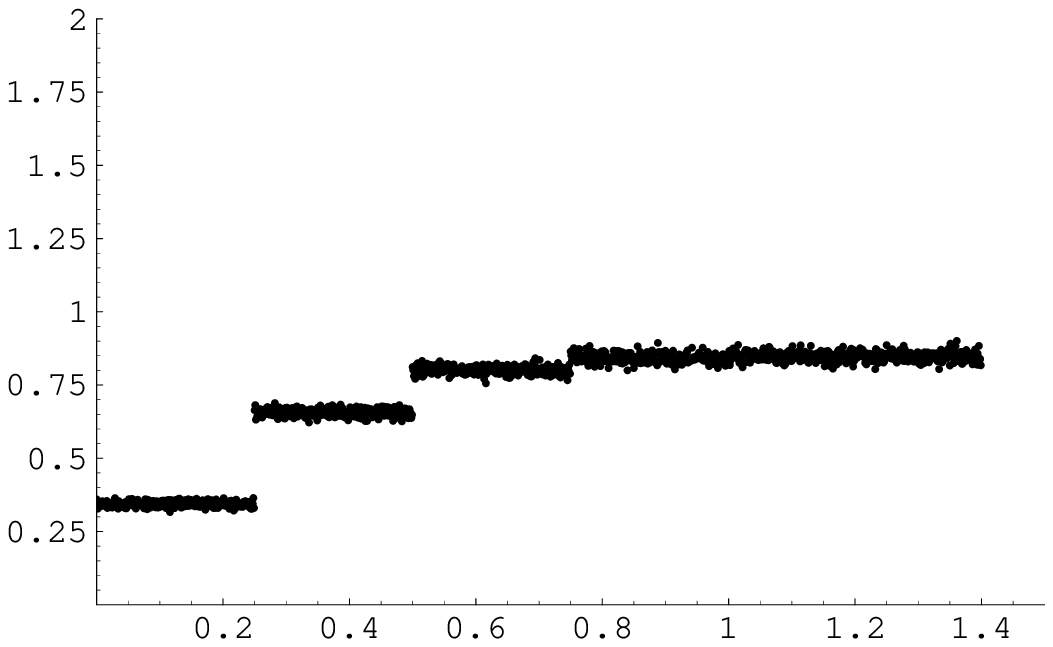}\put(310,-15){(d)}
\end{picture}
\caption{The probability distributions profile for mass different
mass parameters. The mass parameters are $0.05$, $0.1$, $0.2$ and
$0.4$ for the graphs (a), (b), (c) and (d), respectively }
\end{figure}
\begin{eqnarray}
h_u&\rightarrow& h_u-1-t,\nonumber\\
h_{v_i}&\rightarrow& h_{v_i}+1/4,
\end{eqnarray}
where the $v_i$ are the neighbors of the site $u$. As you see in a
toppling process an amount of sand equal to $t$ is lost. One can see
that this model resembles the usual ASM with dissipation: if we take
$t=k/4$, where $k$ is a positive integer number, then the mapping
introduced in previous section, takes this CASM to an ASM which
dissipates $k$ sand grains in each toppling. Simulation results show
that still we have the same plateaus, of course with a new one for
the sites with height greater than one. As we increase the
dissipation, the distances between the plateaus' heights becomes
less, at $t=0.4$ the two last plateaus could not be easily
distinguished. (Fig 3.) The probability distribution function for
the cases $t=0.05,0.1,0.2,0.4$ is shown in Fig. 3.

The other interesting data we can investigate in this model is the
behavior of $p(k)$'s as functions of mass parameter. This has been
done for some WAC's in \cite{MaRu} in an analytical way for
arbitrary value of $t$, though in the original model only positive
integer values of $t$ are meaningful. Again in our model, the mass
could take any positive real number. In Fig. 4 the probability of
finding a site with with different heights is shown and in the case
of height one, is compared with analytic results. The simulation
result coincides with the antilytic one up to $t=0.25$, which
corresponds to one grain of sand loss in ASM. After that, the two
results are not in agreement, we were not been able to find the
origin of this disagreement. Also it is observed that for
probabilities of heights one and two and three, we have a maximum
value that occurs in non vanishing mass parameter, however it seems
that for height four the maximum probabilities happen at $t=0$ or at
a value $t<0.0001$. The maximum values for probabilities of heights
1, 2 and three occur at $t=0.24$, $t=0.08$ and $t=0.004$.

\begin{figure}[b]
\begin{picture}(150,120)(0,-10)
\includegraphics{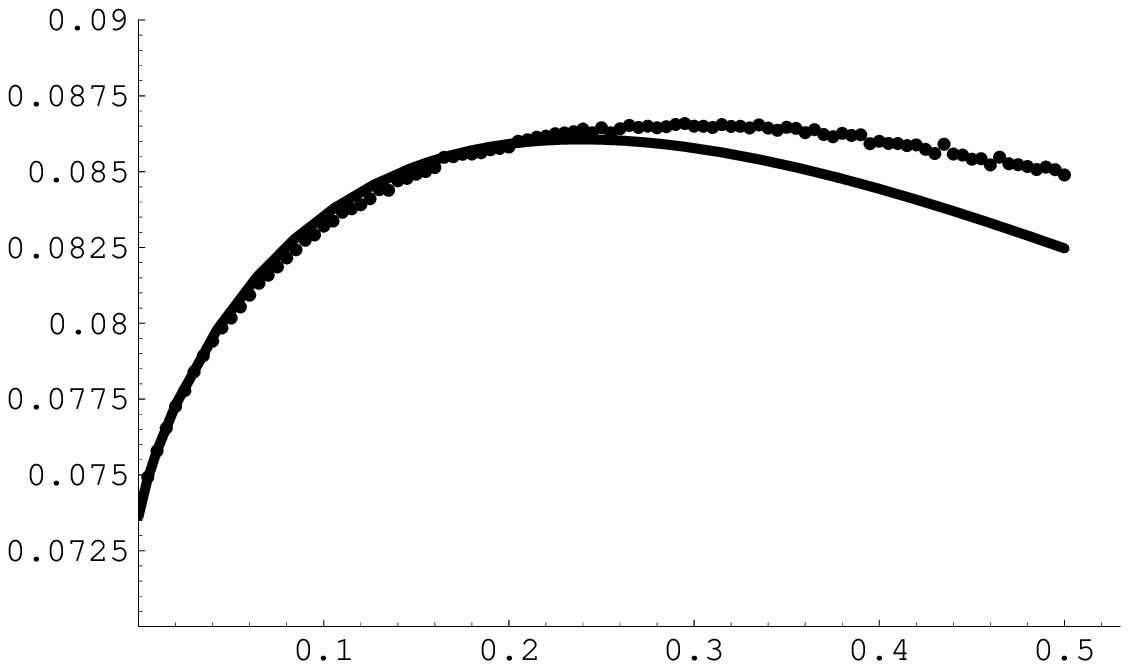}
\includegraphics{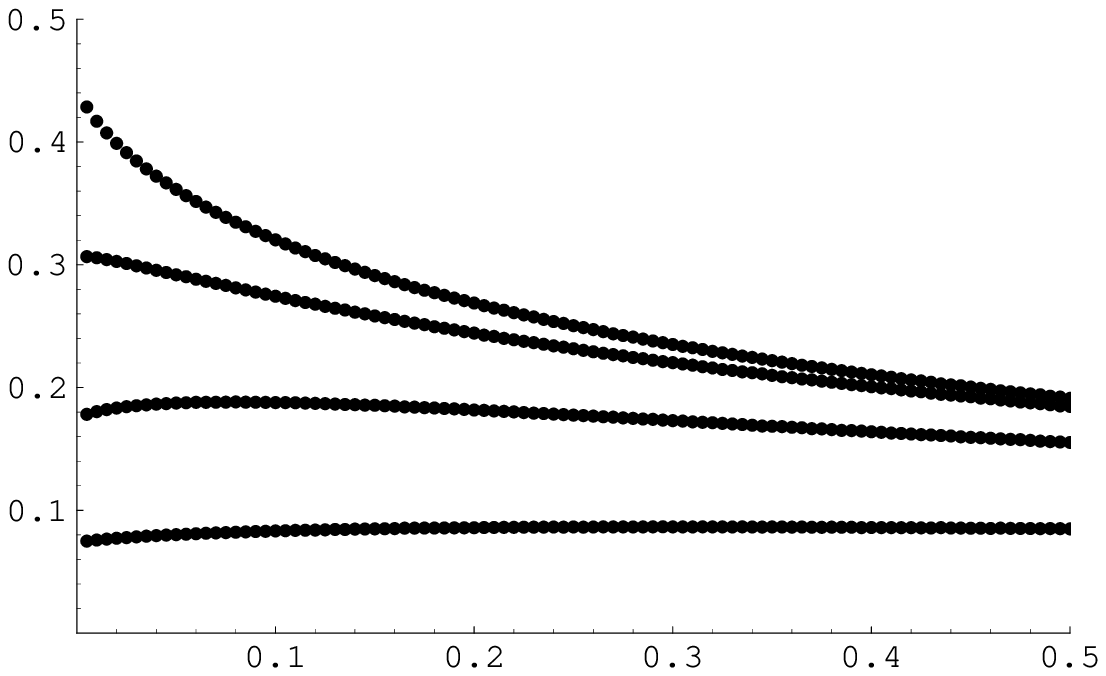} \put(130,-15){(a)}\put(310,-15){(b)}
\end{picture}
\caption{The probability of different heights as functions of mass:
(a) The comparison between simulation and analytical results for
probability of height one. (b) The probabilities of heights 1,2,3
and 4. }
\end{figure}

It is also possible to investigate models where the mass parameter
is not uniform, but is a smooth function of spacial coordinates.
This may lead to some marvelous results if mass is considered as
source of disorder in the theory.

\section{Some Boundary Conditions}
Two well-known boundary conditions (BC's)may be imposed along the
boundaries of usual ASM: open and closed. A boundary site is open if
in a toppling loses 4 grains, one of which falls off the boundary.
It is closed if the number of grains it loses is equal to the number
of its neighbors. Additionally, some other boundary conditions are
investigated \cite{Ruelle07}. Switching from one boundary condition
to other at a specific point, is equivalent to introduce a boundary
changing operator at the point in the field theoretic description of
the model. These operators have been found to have weights equal to
$-1/8$ \cite{Ruelle02,Ruelle07}. In our model, we are able to go
from one boundary condition to the other in a smooth way. The open
boundary site could be thought as a closed boundary site which has a
mass parameter equal to one (in ASM language). Now we can introduce
half-open boundary sites where their mass parameter is nighter zero
nor one, but something in between. Although it is clear that under
renormalization, mass parameter increases and therefore any small
nonzero value of it will eventually have the same effect, but it is
still interesting to examine this. Also we have the opportunity to
alter other elements of the toppling matrix in a continuous way. To
do this, we introduce a new boundary condition called reflective
boundary condition. In the new boundary condition, not only we take
the matrix elements $\Delta_{ii}$ be $3/4 +t$ for the boundary
sites, but we also assume that $\Delta_{ij}=-(1/4 +t)$ for the link
between the boundary sites and the sites who are just one row above
the boundary. This means that there is no dissipation, as one can
verify $\sum_j \Delta_{ij}=0$ for all the sites. There is an
equivalent version of such boundary condition in ASM: if the height
of a boundary site becomes more than four, then it topples, giving
one sand grain to each of its neighbors who sit at the boundary
line, and two sand grains to its other neighbor; that is, instead of
throwing a sand grain away, it is reflected towards the system. The
boundary condition we consider, is a continuous version of such
condition.

Let's first concentrate on open and closed boundary conditions. It
has been shown that operator effecting the change from closed to
open, or from open to closed, is a boundary primary field of weight
-1/8, belonging to a $c = -2$ logarithmic conformal field theory
\cite{Ruelle02}. Such changes happen abruptly, the matrix elements
$\Delta_{ii}$ corresponding to boundary sites change from 3 to 4
just at a single point. We would like to see if these changes are
made smoother, what happens to the boundary operator. As an example
one can take $\Delta_{ii}= 3+(1+ \tanh(\alpha(i-i_0)))/2$ where
$\alpha$ is an arbitrary real number. Although by numerical results
one can see that boundary operators have the same $-1/8$ scaling
dimension, but one is also abe to derive some analytical results.
Let's take the following boundary condition
\begin{eqnarray}
\Delta_{ii}= \left\{
\begin{array}{cc}
1 & i<1\:\: {\rm or}\:\: i>n \\
3/4 &  1<i<n\\
3/4+1/8 & i=1\:\:{\rm and}\:\: i=n
\end{array}
\right.
\end{eqnarray}
This means that the site in the interval $I_0=[2,n-1]$ are closed,
the sites $1$ and $n$ are half open and other boundary sites are
open( See Fig. 5). The ratio of partition functions of theory with
two different boundary condition (e.g. our BC and open BC) yields
the two point correlation function of boundary changing operators.
Following \cite{Ruelle02,Ruelle07} we define a defect matrix $B$
which in our case will be 1/4 of identity matrix in the subspace of
the sites in $I_0$, and has two other nonzero matrix elements
$B_{11}=B_{nn}=1/8$. Therefore the ratio of the partition functions
turns out to be
\begin{figure}
\begin{picture}(80,40)(-40,0)
\includegraphics{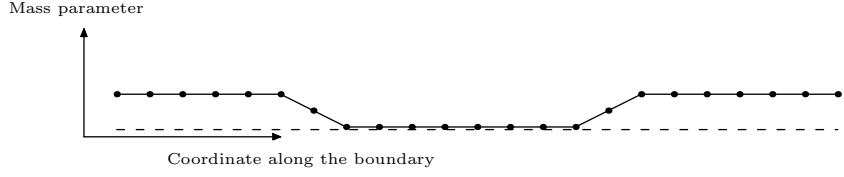}\put(20,57){\tiny Mass parameter}\put(80,0){\tiny
Coordinate along the boundary}
\end{picture}
\caption{A non-sharp boundary condition: Two of the boundary sites
are half-open}
\end{figure}
\begin{equation}\label{determinant2}
\frac{Z^{new}}{Z^{op}}= A
e^{\frac{-2G}{\pi}(n-1)}n^{1/4}\left(\left(1-\frac 2
n\right)^{\frac{1}{4}}e^{\frac{2G}{\pi}}+2\left(1-\frac 1 n
\right)^{\frac{1}{4}}+e^{\frac{-2G}{\pi}}\right)
\end{equation}
Here $A$ is a constant and $G$ is the Catalon constant. As it is
clear, for large $n$'s, we will arrive at the results obtained in
\cite{Ruelle02}. The next leading terms behave as $n^{-3/4}$ that
could be assigned to some operators having scaling dimension equal
to $3/8$. Such fields do exist in $c=-2$ theory.

The other BC we investigate here is the reflecting BC. Let's take
the boundary site $(i,0)$ to be a reflecting one. Let $(i,1)$ be the
bulk site which is a neighbor of the site $(i,0)$. As said before
the reflecting BC implies that $\Delta_{(i,0)(i,0)}=3/4+t$ and
$\Delta_{(i,0)(i,1)}=-1/4-t$. All other matrix elements are
unchanged. Following \cite{Ruelle02} we take a segment of the
boundary with length $n$ to be a reflecting BC, while other sites
are open/closed. The ratio $Z_{R}/Z_{op/cl}$ will give us the two
point function of boundary changing fields. Here $Z_{R}$ is the
partition function of the model with the boundary condition
expressed above and $Z_{op/cl}$ is the partition function of the
model if all the boundary sites are assumed to be open/closed.

All determinants have the Toeplitz form $ \det(\sigma^{\alpha\beta}
_{i-j})$ where $\alpha$ and $\beta$ can take values 0 and 1 and
determine whether we are considering the boundary site or the site
just one row above the boundary. As the matrix elements
$B_{\alpha=1,\beta}=\Delta^{R}_{\,\,\alpha=1,\beta}-\Delta^{cl/op}_{\,\,\alpha=1,\beta}=0$,
the determinants we are considering reduce to determinants of
$n\times n$ matrices. The entries of these matrices are
\begin{eqnarray}
\sigma^{cl/op}_{m}=\delta _{m,0}+\frac {t-a} {4}
G^{cl/op}((0,1);(m,1))- \frac t 4 G^{cl/op}((0,2);(m,1))
\end{eqnarray}
where $a=0,1$ refer to the cases closed/open and $G^{cl/op}$ is
inverse of $\Delta^{op/cl}$. By the method of images one finds that
the coefficients $\sigma_{m}$ are the Fourier coefficients of a
function $\tilde{\sigma}$ ,
\begin{eqnarray}\label{sigmaa}
\tilde{\sigma}^{op}(k)&=&\left(2-2\cos(k)\right)^{\frac{1}{2}}\nonumber\\
&&\hspace{-0.5cm}\left(\left(2t\left(\cos(k)-3\right)+(t-1)\right)\sqrt{\frac{1-\cos(k)}{2}}+\left(2t(2-\cos(k))-(t-1)\right)\sqrt{\frac{3-\cos(k)}{2}}\right)\\
\tilde{\sigma}^{cl}(k)&=&(1+t)+t\left(2-2\cos(k)\right)^{\frac{1}{2}}\left(\sqrt{\frac{1-\cos(k)}{2}}-\sqrt{\frac{3-\cos(k)}{2}}\right).
\end{eqnarray}
The asymptotic value of Toeplitz determinant is given by using a
theorem due to Widom \cite{widom}. Using this theorem one finds
$\langle\phi_{(op,R)} \phi_{(R,op)}(n)\rangle\sim n^{1/4}$ and
$\langle\phi_{(cl,R)} \phi_{(R,cl)}(n)\rangle\sim n^{0}$. Therefore
the scaling exponents of the BC changing operators $\phi_{(op,R)}$
and $\phi_{(cl,R)}$ are $-1/8$ and zero respectively; that is, the
reflecting BC is more or less the same as closed BC, which could be
expected by hand-waving arguments.

At the end we would like to say that some other bulk models could be
developed when the elements of toppling matrix $\Delta$ could be
real values: one can add some spacial asymmetries or add ellipticity
to the problem. This issue is under consideration\cite{AM}.
\vspace{5mm}

{\Large \bf Acknowledgment}

We would like to thank S. Rouhani for his helpful comments and
careful reading of the manuscript.

\end{document}